# Electronic Voting: the Devil is in the Details

Chantal Enguehard
LINA - UMR CNRS 6241
2, rue de la Houssinière, BP 92208, 44322 Nantes Cedex 03, France
chantal.enguehard@univ-nantes.fr

Jean-Didier Graton
Expert in law, technology and economics of IT security, Chairman of the European Computer and Communication Security Institute (ECCSI)
Bastion Tower, 20, Place Champs de Mars, 5 - 1050 Brussels, Belgium
info@naccsi.net

**Abstract:**

Observing electronic voting from an international point of view gives some perspective about its genesis and evolution. An analysis of the voting process through its cultural, ontological, legal and political dimensions explains the difficulty to normalize this process. It appears that international organizations are not capable to properly defend the fundamental rights of the citizens. The approach that was taken when DRE voting computers appeared seems to have reoccured with VVAT voting computers and the european e-poll project.

**Keywords:** VVAT, DRE, voting computers, e-voting, electronic voting, e-poll, electronic voting systems

## I - Time lines on facts

### I.1 - Genesis of electronic voting

The history of electronic voting is short. We intend here to give a quick review about the spread of electronic voting systems of all different flavors (voting computers[1], kiosks[2] and internet voting[3]).

We make a distinction between two types of dissemination of electronic voting systems.

The first type is found in countries which have a strong willingness to use electronic voting systems. These e-voting developers countries encourage companies, that are in most cases local, to produce such systems. The majority of these companies were already involved in security or bank activities and invested in this new market because e-voting technology is simple, easy to produce and offers a significant profit margin. In addition the competition between companies is weak because the market is segmented by the necessity for the e-voting products to conform to the election laws of the country it is developed for.

---

[1] To use a voting computer, each elector must go in his/her usual election poll where s/he can register his/her intent of vote directly on the voting computer. At the end of the election period the voting computer gives the sum of the votes that it computed.
The term "voting computer" will be prefered to "electronic voting machine" because the latter is often abbreviated in "voting machine" which already refers to mechanical voting machine (like the lever machine). Using the same term to refer to two different objects inevitably leads to confusion and should be strongly avoided.

[2] Kiosks allow each voter to process his/her vote from any poll station. Thus, all the kiosks are linked to a central serveur which controls the unicity of each vote, registers the intents of votes, and expresses the results at the end pf the voting period..

[3] An internet voting system is a remote voting system whom functionality are close to kiosks. The main difference is that electors don't from from a controlled place (the poll station), but from any computers that is linked on internet.

We find this situation in countries that make extensive use of electronic voting: Netherlands (with the Dutch company NEDAP), but also in the United States (with the American companies Premier[4] or Election Systems and Software-ES&S), in India (CMC Limited is a subsidiary company to the Indian Tata), in Belgium where the voting system has been developed by local firms and also in Brasil: since 1996[5], this country has been using voting computers, specially developed by two American vendors: Unisys (50,000 computers) and Diebold (350,000 computers). These Brasilan voting computers were designed by the Electoral Supreme Court (Tribunal Superior Eleitoral - TSE).

We find the same situation in countries that use electronic systems on a minor scale: Australia developed its own system eVACS, the Estonia states developed their own internet voting systems.

The second type is found in e-voting adopters countries which decided to use voting systems coming from foreign countries: we find Nedap voting computers in Germany, France, Ireland, Italy and Poland; there are Indian voting computers in Nepal; ES&S voting systems have been used in France and United Kingdom; the Spanish company Scytl exports in Finland, Philippines, Australia, Switzerland; Indra, another Spanish company, supplies France.

**I.2 - Success and failures**

E-voting assessments differs according to their authors.

Electoral official organizations are generally very satisfied with the e-voting systems they decided to use, especially among e-voting developers countries.

The electors opinion is much contrasted. We distinguish three typical groups.

1 - The early adopters group is composed by technophiles who are by definition very enthusiastic about e-voting. They lose their faith in technology when it becomes obvious that voting systems are not more reliable than usual computers.

2 - The sceptical group is quantitatively small. It is a heterogeneous group which encompasses computer scientists, technophobes and security experts. This group's tendency is to doubt whether e-voting is useful and to question officials.

3 - The median group brings together the majority of the voters. These persons obviously respect official decisions and, at first, accept the idea that e-voting is a neutral media to vote, even if they observe some practical problems when they vote. By practising e-voting, their opinion moves, generally by analogy with well-known objects or situation (pocket calculators, computers, money-transfer, etc.) that are commonly presented by media, even if they are not appropriate. After an astonishment period, they become capable to evaluate the advantages and disadvantages.

These different populations are confronted by difficulties that appeared after a few elections which used electronic systems.

Many errors were reported in the registration of votes. It was also reported that some voting computers failed to start up, encountered breakdowns or showed some strange behaviour.

For instance, in the USA, the Montgomery County Election Board published a report about the 2004 presidential election which stated that "189 voting units (7%) of units deployed failed on Election Day. An additional 122 voting units (or 5%) were suspect based on number of votes captured" [Montgomery County Election Board 2005]. "A few machines in Miami-Dade County

---

4  Formerly Diebold
5  Electronic voting was generalized through the whole country in 2002.

reset themselves while voters were trying to vote. Precincts in Palm Beach County reported problems activating some of the electronic cards used to authenticate the voters. Even mark-sense ballots designed to be read by optical scanners proved troublesome" [Mercuri 2002]. "In November 2003 in Boone County, Indiana over 144,000 votes were cast even though Boone County contains fewer than 19,000 registered voters." [Simons 2004]

In Belgium, during the 18 May 2003 election day, an error of 4096 votes has been discovered. It had been impossible to explain or reproduce this error, even after examining the faulty computer Rapport Chambre et Sénat belge 2004, page 21].

In Québec, where electronic voting was used to process 95% of the votes in the local elections of 2005, so many problems were encountered that the city decided to return to traditionnal paper ballot: results arrived hours in late, some materials broke down, internet connexion had been cut and many votes had been counted twice by error [Beaulieu 2006].

In addition, researchers demonstrated that some voters were not able to vote independently because of their lack of experience with computers (digital gap) [Birdsall et al. 2005]. It was also discovered that electronic computers were not well adapted to blind people. In addition, controversy arose around the lack of transparency, security, and reliability. Researchers demonstrated that it was impossible to check the results which are automatically delivered and also that results could be tampered with without detection.

With the exception of Ireland, which decided to not use the 75,000 voting computers that were previously bought, most of the concerned states that were deeply involved in electronic voting reacted by asking for improvements, thinking that problems were due to the lack of experience and could be easily fixed. In the USA, for example, with the Help America Vote Act, the congress allocated $3,8 billion to states for improvement of voting system infrastructure, voter education, and training of election officials [Shelley 2004].

In addition, some countries, like Venezuela, moved to the new concept of Voter Verified Audit Trail first expressed by Rebecca Mercuri [Mercuri 2000]. The USA, the South Korea and the Kingdom of Belgium would appear to be following this trend, forgetting the fact that adding complexity to a system always makes it become more fragile and less reliable.

Among the countries that experimented on a very limited scale, some just stopped their trials (Italy, Spain), waiting for a better technology, while some others are still continuing with experiments but without any extension, (France, UK, Switzerland, etc. ).

At a supranational level, Europe goes further by starting deployment of the e-poll system which is supposed to manage "polling preparation, voters' identification and authentication, authorization, vote casting, vote counting and communication of the results" in the name of the reinforcement of the European integration.This projects includes biometric identification, centralization of the uniqueness of each vote and the possibility for the voter to vote from any poll station.

### I.3 - International Organizations

During this period different international organizations, such as the Office for Democratic Institutions and Human Rights of the Organization for Security and Cooperation in Europe (OSCE/ODIHR) or the European Union (EU), sent assessment missions, observing electronic voting systems as traditional voting ones.

Some international organizations produced some guidelines about elections that explicitly take into account electronic voting ([United Nations 2005], [ComVe 2002]).

It seems that international organizations consider electronic voting as a tool which may aid in making elections more peaceful. For many years, these organizations have been confronted with major disorders in polling stations, including voter intimidation, ballot stuffing, ballot box replacement, etc. Voting computers would seem to be a radical solution for such problems[6].

## II - Obstacles to a supranational organization of vote

Many obstacles present themselves when an attempt is made to define a unique set of precise rules which may be applied to the organization of a democratic election, or a unique voting system that could be used by several different countries because an electronic voting system is a highly complex, multi-dimensional object.

**II.1 - Cultural Obstacles**

Traditions are different in different countries, including differences in voting activities. These voting traditions include, for example:

- the format of the ballot: while many countries use an Australian ballot, a French voter still votes by choosing one ballot which holds the name of the candidate or the list that s/he prefers. Australian ballots can present also different aspects: people have to write numbers or to fill a circle or to mark a column, etc. Any change in these habits can have huge consequences and may change the result of the vote[7].

- the susceptibility to fraud depends strongly on the history of the country. Italy encounters many problems because of the informal mafia; in France, before the obligation to use transparent urns, ballot stuffing was common. Young democracies with a little experience of voting procedures are not sufficiently aware of these possibilities for fraud and are therefore unable to take measures to minimize them.

- different vote couting systems are in use: For example Run-Off Voting is used in France, Ireland uses Single Transferable Vote, Germany uses Mixed Member Proportional, Norwegians vote with an Additional Member System, etc.

- in addition the organization of the polling office can take many different forms depending on the size of the polls, the number of urns, the number of officials, the possibility for the voters to actively participate in the organization of the polling day (in France, in each poll office, voters usually count the votes at the end of the poll day), the counting process, the totalization process, the publication of the results, etc.

Differences may exist even within a single country: in France the publication of the results is very different from a city to another. Some make public the detailed results (poll by poll) including the numbers of procurations and signatures on the registry to control the uniqueness of each vote, some others publish only the number or the percentage of the votes obtained by each candidate.

The introduction of a new voting procedure should be prepared by an apprenticeship based on the practical knowledge people get, and targetted to the different categories of population that have to manage with the new system: citizens, electoral organizations, judges, lawyers, etc. The actual lack of knowledge makes the people worrying about false problems. but incapable of detecting and diagnozing eventual real malfunctioning. This situation is very risky because if a real problem occurs and is not detected in time, the consequences could be huge, technically (if ballots are spoiled for instance), thus politically, considering that an election can not be done again.

---

6  This argument is explicitly used by suppliers to convince their customers. On the web site of the manufacturer Unisys which equipped Brazil, we can read a declaration of Paulo Cesar Camarão, Information Technology Director of the Electoral Supreme Court, saying "One of the principal benefits of the solution is that it has completely eliminated the chance for fraud in our elections "

7  In 2007, in Scotland, local elections and scottish parliament election took pace the same day). The usual ballot format has been adapted to this double election, this new ballot caused confusion for several thousands of people whose ballots had been invalidated.

## II.2 - Ontological Obstacles

Eliminating human involvement in the voting tabulation appears to be a protection against fraud, errors or misbehaviour. The human risk is then replaced by technological risk that can be statistically evaluated. This approach is usual in many areas including technological innovations like fast trains, nuclear power plant, etc.

But, in this analysis, the complexity of electronic voting systems has been underestimated because two major dimensions of this object were ignored. Firstly, such systems are supported by computers which are not stable. There are constant updating and upgrading (on software, drivers, periphericals) that make it almost impossible to follow the classical certification approach because a certified object should be recertified each time it is modified. Secondly, the laws that decide the rules for voting are also susceptible to change, forcing the voting system to be quickly adapted to the new law.

This potential constant evolution (electronic and legal) of an electronic voting system constitutes an high level of risk that is not met by any other technological object.

## II.3 - Legal Obstacles

The common voting criteria that are defined by the Venice Commission are very poor and do not fix all the parameters of the voting process. The e-poll project cites this six principles "of Europe's electoral heritage": Universal suffrage, Equal suffrage, Free suffrage, Secret suffrage, Direct suffrage, Frequency of election. These criteria are far from being sufficent to define the conditions that must be in place in the organization of a democratic election: there is no word about transparency and the necessity for the voters themselves to control the election and eventually contest the results [Enguehard 2008].

Elections depend on the respect of fundamental values which have not been enough precisely defined in Europe. The European Convention on Human Rights defines the right to vote for every citizen but does not interfere in the electoral contentious affairs that could lead to invalidate an election. This text is only a framework and does not give any information about how to conduct an election. Therefore the European Court on Human Rights is not competent on elections.

The Venice Commission does not define explicitly the essential independance of the organisationnal body in the case of electronic voting. This question is crucial because it is impossible to check whether a computer is independent or not.
In addition this commission confused mechanical and electronic voting methods and thus produced recommandations on e-voting that are not operational. For example, the principle 43 states "Electronic voting methods must be secure and reliable" and gives some definition of reliability and security but do not precise how the level of security and reliability should be evaluated. The principale 44 adds that "the system's transparency must be guaranteed in the sense that it must be possible to check that it is functioning properly." but the range of this transparency and the conditions to get an effective transparency are not enonciated.

## II.4 - Political Obstacles

Introducing electronic voting causes deep changes in the habits and rights of the citizens who may, therefore, be unwilling to accept this evolution. In France, in addition to the officials, traditional polls are under the eyes of scrutinizers, official delegates from parties and magistrats, the counting is directly processed by citizens. The introduction of electronic voting simply suppressed scrutinizers while officials, official delegates from parties and magistrats can no longer control anything except what the voting computer accepts to display or print. This lack of popular, political and legal invigilation contributes to diminish the confidence in the electoral system.

We have also to note the resurgence of regionalism in Europe, since the 1980s which is a political translation of the cultural differences [Keating 2003].

## III - Limits

Many international assessment missions took place in countries where electronic voting systems are in use: DRE voting systems[8], VVAT voting systems[9], or internet voting systems[10].

### III.1 - Limited observations on DREs

Several observation mission did took place in countries using DRE voting machines. The lack of transparency of these new technological systems did not alert the international observers at the first stage because their guidelines did not include these new objects.

While UEM do not say a word about electronic voting in its observation guidelines, the ODIHR/OSCE updated its Election Observation Book, in an effort to take in account the apparition of electronic voting in its fifth edition in 2005.

Becoming conscious of the inobservability of DRE voting computers, the ODIHR/OSCE sent an expert mission in 2006 in Belgium "to increase [its] comparative knowledge of e-voting systems, also with a perspective on how to most effectively observe such processes. " This report express clearly that "observation of the e-voting system is *de facto* limited to an analysis of the security mechanisms in place, and to an observation of their implementation." [odihr/OSCE Belgium 2006].

### III.2 - A case studies: the Voter-Verified Audit Trail (VVAT)

A VVAT voting computer presents to each voter a printed version of its ballot in order to give him/her the possibility to check that it complies with his/her choices. These ballot papers are kept in a urn, so there is a theoritical possibility to verify the results of the computers by counting manually the content of the urn. The main idea is that only a part of the urns would have to be recounted (to save time and money).

Many international organizations produces documents that tend to give favor to VVAT voting computers.

The Venice Commission[11] expressed "electronic voting should be used only if it is safe and reliable; in particular, voters should be able to obtain a confirmation of their votes and to correct them, if necessary, respecting secret suffrage; the system must be transparent;" but it did not give any operational definition of transparency to define what should be transparent to who.

The ODIHR/OSCE stated "In the absence of a paper trail, which could allow the voters to verify the accuracy of their vote, and would provide for possibilities of a paper recount in case of doubt, there is no way the above mentioned aspects can be directly observed."[odihr/OSCE 2006]

The ODIHR/OSCE observation book simply warns to be aware of "Electronic voting systems with no voter-verified auditable paper trail or other manual audit capacity.", letting think that electronic voting systems with audit capacity are acceptable.[odihr/OSCE 2005]

All these guidelines have been conceived in reaction to the opaqueness of the DRE voting computers and have been built on a consensus between the participants without working towards a complete analysis of the new system they favor, following the same approach that let the DRE systems been deployed without any reaction.

---

8  Presidential election in USA [odihr/OSCE USA 2004], federal election in Belgium [odihr/OSCE Belgium 2007], presidential election in France [odihr/OSCE France 2007], parliamentary elections in Netherlands [odihr/OSCE Netherlands 2007]
9  Parliamentary elections [EU EOM Venezuela 2005] and presidential elections in Venezuela [EU EOM Venezuela 2006]
10 Parliamentary elections in Estonia ([odihr/OSCE Estonia 2007]
11 which confuse mechanical and electronic voting methods, ignoring the ontological difference of these two objects.

An analysis of The VVAT reveals that there is a complete lack of safeguards to frame this recount possibilities: the VVAT process is verifiable that is far to mean verified. Technically, if its results are not verified a VVAT is equivalent to a DRE voting computer, including all the DRE defaults that are now well known[12].

So, VVAT voting computers pretend to improve the reliability of the voting process. Actually, if there is no recount, there is no reliability. At least the right to recount should be unforce by law.

Thus, observations must precisely concentrate their attention on the real verifications that are made and they must deal with complex questions that imply legal, organizational and technical dimensions. For instance, it should be decided what should happen, if a voter claims that the printed ballot paper does not reflect her/his choice (will s/he be believed, being unable to prove her/his intent?). How, when and by who are chosen the computers whose results are audited? Who do recount manually the ballot papers? Are the ballot papers counted just after the election and in the poll stations, or days after election in another place? Who can ask for a recount and obtain it? Legally, what should happened if the ballots counting differs from the automatic counting? etc.

These questions are not yet taken in account as is evident from the observation mission that was deployed for the parliamentary election in Venezuela. VVAT computers were used to vote but the observation mission did not pay enough attention to these questions [EU EOM Venezuela 2005]. The second mission, to a presidential election, just noted a few percentages relating to the increased number of polling stations in which the voting receipts were counted but were not present to observe whether the choice of the machine was really random or whether best security practices were adhered to during the transportation and recount of the ballots [EU EOM Venezuela 2006].

Finally it should be kept in mind that even if the right to recount becomes guaranteed, real recounts will become more and more rare because, historically, we can see that once a task is done by computers, humans cease to manage it. VVAT computers have been build to conquest the lost confidence of the voters. Voters will ask for recounts until they recover their confidence in the voting system, and then, once the confidence back, they will stop to ask for recounts.

**III.3 - A case studies: the e-poll project**

The e-poll project was one of the e-en project founded by Europe. This project is based on the Venice Commission which reflects the consensus between the partipant countries. We did not find any deep analysis about the feasability of this project. It is a kiosk system, allowing people to vote from any poll station. Biometry is used to identify the voter. The system prints a receipt to give the voter an insurance that his/her vote has been taken in account.

Experian, the company responsible for the "Regulations and Specifications" working package had several experiments in electronic voting. It organized the vote for the French living abroad in 2006. One of the reports, published by an expert who audited the system during the vote period, revealed that there were no safeguard to close poll stations where only a very few voters (one or two) had been previously registered. These poll stations opened despite the risk of not respecting the confidentiality of the cast vote [Pellegrini 2006]. In fact , during the vote period, Experian suddenly became aware of this problem and decided to close several poll stations with a few registered voters (Kabul, Bandaar, Riga, Skopje Tbilissi, Reykjavik, Colombo, Ulan-Bator, etc.), destroying the few votes that were already cast. Voters were not informed that their votes had been canceled in this operation.

The option of voting from any polling station, which electronic voting systems may offer, can lead to similar problems. If all the countries in Europe are taken into account it is strong possibility that a citizen of a country with a small population might well be the only voter taking part in his national elections from a particular foreign polling station.

---

12 A VVAT computer is a DRE computer with a printer and a process to conduct the ballot paper to the urn. This additional features may breakdown and thus, fragilize the entire process.

A court case revealed that during the same election dozens of citizens voted with the same computer. The decision makes clear that the judge did not realize that this "group voting" reveals a loss of confidentiality [Conseil d'État 2007].

This case is a simple example which shows that the poor capability to anticipate problems is often a characteristic of e-voting implementations.

In addition, the capacity to identify the voter, register its vote and respect the confidentiality of the vote cannot be demonstated to voters. Generally speaking, this project does not consider the crucial dimension of transparency, despite its strong involvement in legitimating the results of the election process.

Finally, e-poll claims that it is "fostering the active participation of the population in the voting process", whereas all the experience of the last 15 years shows that electronic voting does not increase the turnout rate. Actually, opposite to an "active participation", this system deprives citizens of the right to control the voting process, forcing them to a passive attitude.

This project presents has many other deficiencies that can not be detailed here.

## Conclusion

Many governments which face a strong political opposition are waiting for the international organizations to standardize voting procedures and electronic voting systems, even if several scientists claim that the defined standards present serious problems and that such certification procedures do not solve the majority of security or usability problems [Alexander 2004], [Mulligan 2004], [McGaley 2006], [Barr 2007]. Until now, international organizations limit themselves to a search for political consensual positions which make up only a minimal set of the conditions that are necessary to ensure democratic elections. In addition their assessment missions are limited to observing only what is defined in this incomplete consensus, while ignoring more pertinent criteria.

With the case of the e-poll project we saw how Europe fails to protect the citizens' interests because of the lack of definition of a precise set of fundamental rights. Such a goal can be attained only through negotiation because some countries would have to change to conform the new rules, and not by consensus.

The economic dimension is also at stake because electronic voting is now a market whose interests may not be the same than the citizens' ones.